\documentclass{ws-procs11x85}
\def\GeV{{\rm GeV}}

\makeindex
\begin{document}

\title{Parton Distributions}

\author{R. S. THORNE}

\address{Cavendish Laboratory, University of Cambridge, \\
Madingley Road, Cambridge, CB3 0HE, UK\\E-mail: thorne@hep.phy.cam.ac.uk}


\twocolumn[\maketitle\abstract{I discuss our current understanding of parton 
distributions. I begin with the underlying theoretical framework, and the way 
in which different data sets constrain different partons, highlighting 
recent developments. The methods of examining the uncertainties 
on the distributions 
and those physical quantities dependent on them is analysed. Finally I look at
the evidence that additional theoretical corrections beyond NLO perturbative
QCD may be necessary, what type of corrections are indicated and the impact 
these may have on the uncertainties.}]

\baselineskip=13.07pt

\section{Introduction}

The proton is described by QCD -- the theory of the strong 
interactions. This makes an understanding of its structure a difficult problem.
However, it is also a very important problem --
not only as a question in itself, but also in order to search for and 
understand new physics. 
Many important particle colliders use hadrons -- HERA
is an $ep$ collider, the Tevatron is a $p \bar p$ 
collider, the LHC at CERN will be a $pp$ 
collider, and an understanding of proton structure is essential in order to
interpret the results. Fortunately, 
when one has a relatively large scale in the process,
in practice only $> 1 \GeV^2$, 
the proton is essentially made up of the more fundamental
constituents -- quarks and gluons (partons), which interact relatively weakly.
Hence, the fundamental quantities 
one requires in the calculation of scattering 
processes involving hadronic particles 
are the parton distributions.
These can be derived from, and then used within, 
the {\it factorization theorem} which separates 
processes into nonperturbative parts which can be determined from 
experiment,
and perturbative parts which can be calculated as a power-series in the 
strong coupling constant $\alpha_S$.

This is illustrated in the canonical example of deep inelastic scattering. 
The cross-section for the virtual photon-proton interaction 
can be written in the factorized form
$$
\sigma(ep \to eX) = \sum_i C^{DIS}_{i}(x,\alpha_s(Q^2))\otimes
f_{i}(x,Q^2) 
$$
where $Q^2$ is the photon virtuality, $x=\frac{Q^2}{2m\nu}$, the momentum 
fraction of parton ($\nu$=energy transfer in the lab frame), and the 
$f_{i}(x,Q^2)$ 
are the parton distributions, i.e the 
probability of finding a parton of type $i$ carrying a 
fraction $x$ 
of the momentum of the hadron. 
Corrections to the above formula are of  
${\cal O}(\Lambda_{{\rm QCD}}^2/Q^2)$ and are known as 
higher twist. The parton distributions are 
not easily calculable from first principles. However, they do evolve with 
$Q^2$ in a perturbative manner, satisfying the evolution equation
$$ 
\frac{d f_{i}(x,Q^2)}{d \ln Q^2} = \sum_i 
P_{ij}(x,\alpha_s(Q^2))\otimes f_{j}(x,Q^2) 
$$
where the splitting functions $P_{ij}(x,\alpha_s(Q^2))$
are calculable order by order in 
perturbation theory.
The coefficient functions $C^P_{i}(x,\alpha_s(Q^2))$ 
describing a hard 
scattering process are process dependent but are calculable as a 
power-series, i.e
$C^P_{i}(x,\alpha_s(Q^2))= \sum_k C^{P,k}_{i}(x)\alpha^k_s(Q^2).$
Since the $f_{i}(x,Q^2)$ 
are process-independent, i.e. {\it universal}, once they have been measured at 
one experiment, one can predict many other scattering processes.

Global fits\cite{MRST2001}$^-$\cite{ZEUSfit} use all available data, 
largely structure functions, and the most 
up-to-date QCD calculations, currently NLO--in--$\alpha_s(Q^2)$,  
to best determine these parton distributions and
their consequences. In the global fits input partons are parameterized as, e.g.
$$xf(x, Q_0^2) = (1-x)^{\eta}(1+\epsilon x^{0.5}+\gamma x)
x^{\delta}$$
at some low scale 
$Q_0^2 \sim 1-5 \GeV^2$, and evolved upwards using NLO
evolution equations. Perturbation theory should be valid if 
$Q^2 > 2 {\rm GeV^2}$, and hence one fits data
for scales above $2-5 \GeV^2$, and this cut should also remove the 
influence of 
higher twists, i.e. power-suppressed contributions. 

In principle there are many different parton distributions -- all quarks and
antiquarks and the gluons. However, $m_c, m_b \gg 
\Lambda_{{\rm QCD}}$ (and top does not usually contribute), 
so the heavy parton distributions are 
determined perturbatively. Also we usually assume $s=\bar s$, and that
isospin symmetry holds, i.e. $p \to n$ leads to $ d(x) \to u(x)$ and 
$u(x) \to d(x)$. This leaves 6 independent 
combinations. Relating $s$ to $1/2(\bar u + \bar d)$ we have the 
independent distributions
$$u_V = u- \bar u, \, d_V =d-\bar d, 
\, {\rm sea}=2*(\bar u + \bar d + 
\bar s), \, \bar d - \bar u, \, g. $$
It is also convenient to define 
$ \Sigma = u_V + d_V + {\rm sea} +(c+\bar c) +(b+\bar b)$.
There are then various sum rules constraining parton inputs and 
which are conserved by evolution order by order in 
$\alpha_S$, i.e. the number of up and down valence quarks   
and the momentum carried by partons (the latter being 
an important constraint on the gluon which is only probed indirectly), 
$$ \int_0^1 x\Sigma(x) +x g(x) \, dx =1. $$

When extracting partons one needs to consider that not only are there 
6 independent 
combinations, but there is also a wide distribution of
$x$ from $0.75$ to $0.00003$. 
One needs many different types of experiment for
a full determination.   
The sets of data usually used are: 
H1 and ZEUS $F^p_2(x,Q^2)$ data\cite{H1A,ZEUS} which covers 
small $x$ and a wide range of $Q^2$; E665 
$F^{p,d}_2(x,Q^2)$ data\cite{E665} at medium $x$;
BCDMS and SLAC $F^{p,d}_2(x,Q^2)$ data\cite{BCDMS,SLAC} at large $x$; 
NMC $F^{p,d}_2(x,Q^2)$\cite{NMC} at medium and large $x$; 
CCFR $F^{\nu(\bar\nu) p}_2(x,Q^2)$ and
$F^{\nu(\bar\nu) p}_3(x,Q^2)$ data\cite{CCFR} at large $x$ 
which probe the singlet and valence quarks independently;
ZEUS  and H1 $F^{p}_{2,charm}(x,Q^2)$ data\cite{ZEUSc,H1c}; 
E605 $ p N \to \mu \bar \mu + X$\cite{E605} constraining the large $x$ 
sea; E866 Drell-Yan asymmetry\cite{E866} which determines  
$\bar d -\bar u$; CDF W-asymmetry data\cite{Wasymm} 
which constrains the $u/d$ ratio at 
large $x$; CDF and D0 inclusive jet data\cite{D0,CDF} 
which tie down the high $x$ gluon; 
and NuTev Dimuon data\cite{NuTeV}
which constrain the strange sea.

The determination of the different partons in given kinematic ranges can be 
split into a few different classes. We begin with 
{\bf large $\bf x$}. Here the quark distributions    
are determined mainly from structure 
functions, which are dominated by non-singlet valence distributions.  
Both the evolution of these non-singlet distributions and 
conversion to structure functions is quite simple involving no parton 
mixing 
\begin{eqnarray} \frac{d f^{NS}(x,Q^2)}{d \ln Q^2} &=&  
P^{NS}(x,\alpha_s(Q^2))\otimes f^{NS}(x,Q^2) \nonumber \\
F_2^{NS}(x,Q^2)& = & C^{NS}(x,\alpha_s(Q^2))\otimes
f^{NS}(x,Q^2). \nonumber \end{eqnarray}
Hence, the evolution of high $x$ structure functions 
is a good test of the theory and of $\alpha_S(Q^2)$. The success is shown 
in Fig.~1. 
However - perturbation theory involves contributions 
to the coefficient functions
$\sim \alpha_S^n(Q^2) \ln^{2n-1}(1-x)$ 
and higher twist contributions are known to be enhanced 
as $x \to 1$. Hence, in order to to avoid contamination of NLO theory
one makes a cut $W^2 = Q^2(1/x-1)+m_p^2 \leq 10-15 \GeV^2.$

\begin{figure}[h!]
\vspace{-1.5cm}
\begin{center}
\psfig{figure=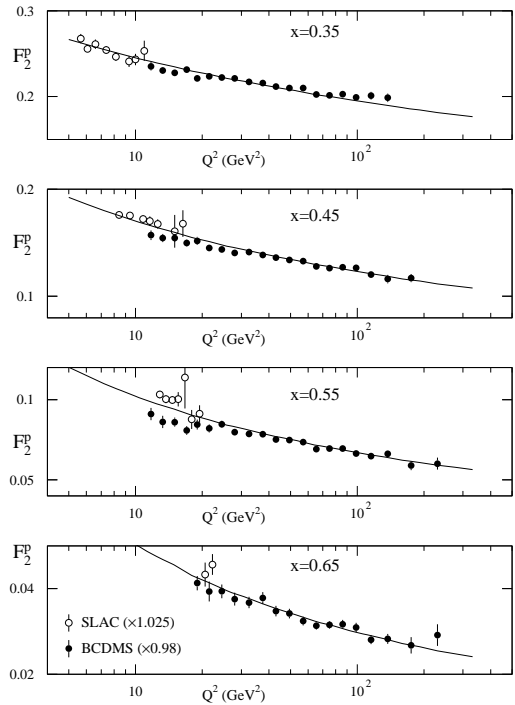,width=7.5truecm}
\end{center}
\vspace{-0.6cm}
\caption[]{Description of large $x$ 
BCDMS and SLAC measurements of $F_2^p$.}
\label{largex}
\end{figure}

\vspace{-0.4cm}

The extension to very {\bf small $\bf x$} 
has been made in the past decade by HERA.
In this region there is very great scaling violation of the partons from the
evolution equations and also interplay between the quarks and gluons. 
At each subsequent order in 
$\alpha_S$ each splitting 
function and coefficient function obtains an extra power of $\ln(1/x)$ 
(some accidental zeros in $P_{gg}$), i.e. 
$ P_{ij}(x,\alpha_s(Q^2)),C^P_i(x,\alpha_s(Q^2)) \sim 
\alpha_s^m(Q^2)\ln^{m-1}(1/x),$
and hence the convergence at small $x$ is questionable.
The global fits usually assume that this turns out to be 
unimportant in practice, and proceed regardless. The fit is good, but 
could be improved.
The large $\ln(1/x)$ terms mean that 
small $x$ predictions are somewhat uncertain, 
as will be discussed later.
Small $x$ parton distributions are therefore an interesting field of study 
within QCD. 
They are also vital for understanding the standard production 
processes at the LHC, and perhaps some of the more exotic ones,
as shown in Fig.~2, which demonstrates the range of $x$ 
probed by the experiment. 

\vspace{0.8cm}

\begin{figure}[h!]
\begin{center}
\vspace{-0.0cm}
\psfig{figure=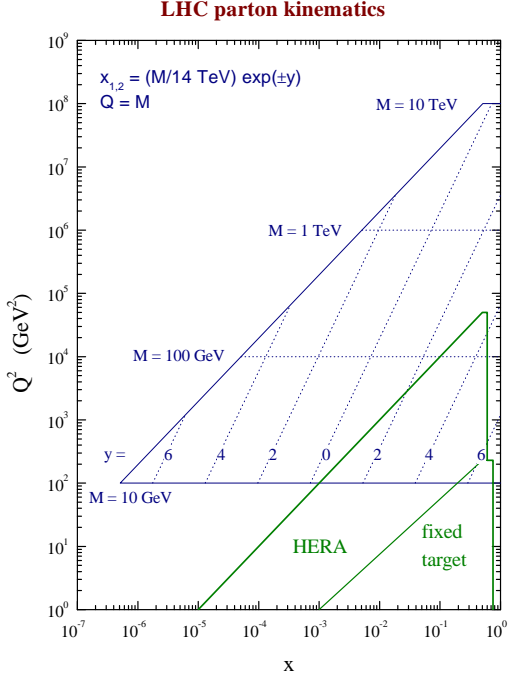,width=7.5truecm}
\end{center}
\vspace{-0.5cm}
\caption[]{The range of $x$ probed at HERA and the LHC as a function of the 
hard scale, e.g. particle mass, and rapidity.}
\label{LHCkin}
\end{figure}

\vspace{-0.5cm}

\begin{figure}[h!]
\begin{center}
\psfig{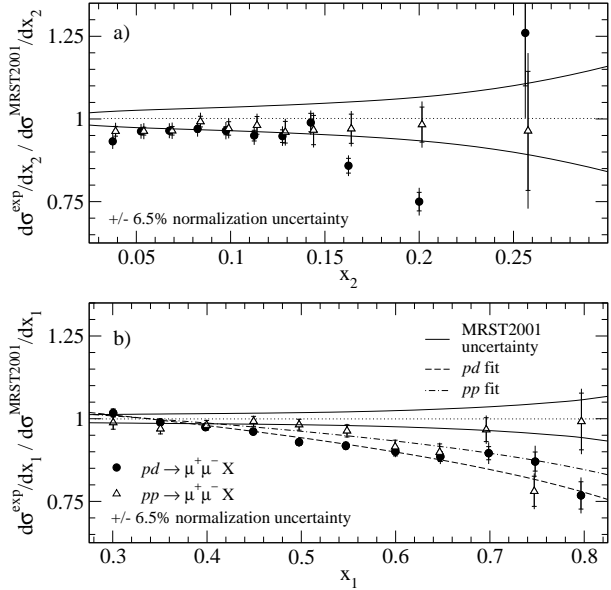}
\end{center}
\vspace{-0.2cm}
\caption[]{E866 fit to their Drell-Yan data as a function of $x_1$ (quark) 
and $x_2$ (antiquark).}
\label{E8661}
\end{figure}

\vspace{-0.1cm}

The {\bf high-$\bf x$ sea quarks} are determined by Drell-Yan data
(assuming good knowledge of the valence quarks).
There is new precise data from the E866/NuSea 
collaboration\cite{E866/NuSea}, 
and their fit to these data shows a discrepancy with existing partons
implying larger high-$x$ valence quarks, as shown in Fig.~4.
However, the fit performed by MRST (Fig.~4) and 
CTEQ displays no such discrepancy.

\vspace{-0.8cm}

\begin{figure}[h!]
\begin{center}
\includegraphics[width=.47\textwidth]{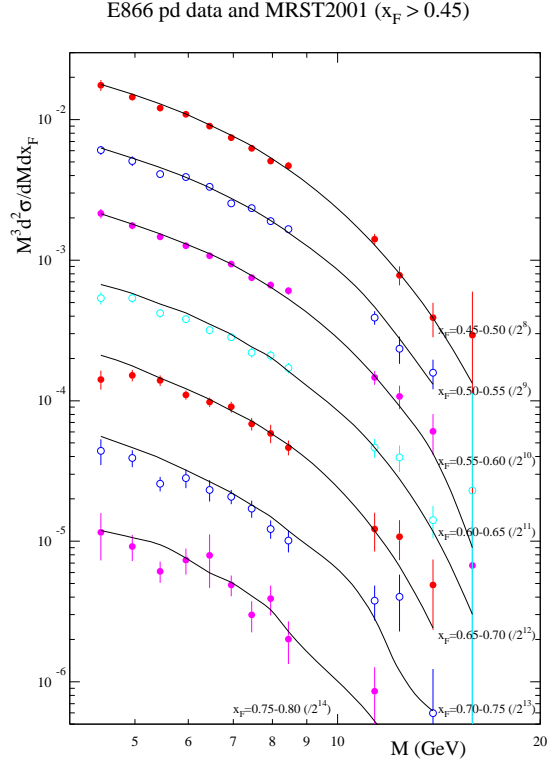}
\end{center}
\vspace{-0.4cm}
\caption[]{MRST fit to the E866 data for high values of $x_F$.}
\label{E8662}
\end{figure}

The $\bf s(x)$ {\bf and $\bf \bar s(x)$ distributions}
are probed using CCFR and NuTeV dimuon data, i.e. the processes  
$$
\nu + s \to \mu^- + c (\mu^+), \quad
 \bar \nu + \bar s \to \mu^+ + \bar c (\mu^-).
$$
The quality of data is now such that one can examine the $s(x)$ and 
$\bar s(x)$ distributions
separately. This has recently been performed in detail by CTEQ\cite{CTEQs}. 
They find that $s(x) < \bar s(x)$ at quite small $x$, but since
$\int (s(x)-\bar s(x))\,dx=0$, (zero strangeness number) this leads to 
$\to \int x(s(x)-\bar s(x))\,dx =[S^-]>0$, as demonstrated in Fig.~5. 
They obtain the rough constraint $0 < [S^-] <0.004$. 
This is particularly significant because 
$NuTeV$ measure\cite{zeller} 
$$
R^-=\frac{\sigma^{\nu}_{\rm NC}
-\sigma^{\bar\nu}_{\rm NC}}{\sigma^{\nu}_{\rm CC}
-\sigma^{\bar\nu}_{\rm CC}},
$$
and in the standard model this satisfies
$R^-=\frac{1}{2}-\sin^2 \theta_W -(1-\frac{7}{3}
\sin^2 \theta_W) \frac{[S^-]}{[V^-]}$. There is currently a $3\sigma$
discrepancy between this determination of $\sin^2 \theta_W$ and 
others\cite{gambino} but  
$[S^-] =0.002$ reduces this anomaly from $3 \sigma$ 
to $1.5\sigma$. NuTeV themselves claim no such strange asymmetry when 
using partons obtained from fitting their own data\cite{NuTeV}, 
so this is an issue which requires resolution. 

\begin{figure}
\hspace{0.05cm}
\includegraphics[width=.46\textwidth]{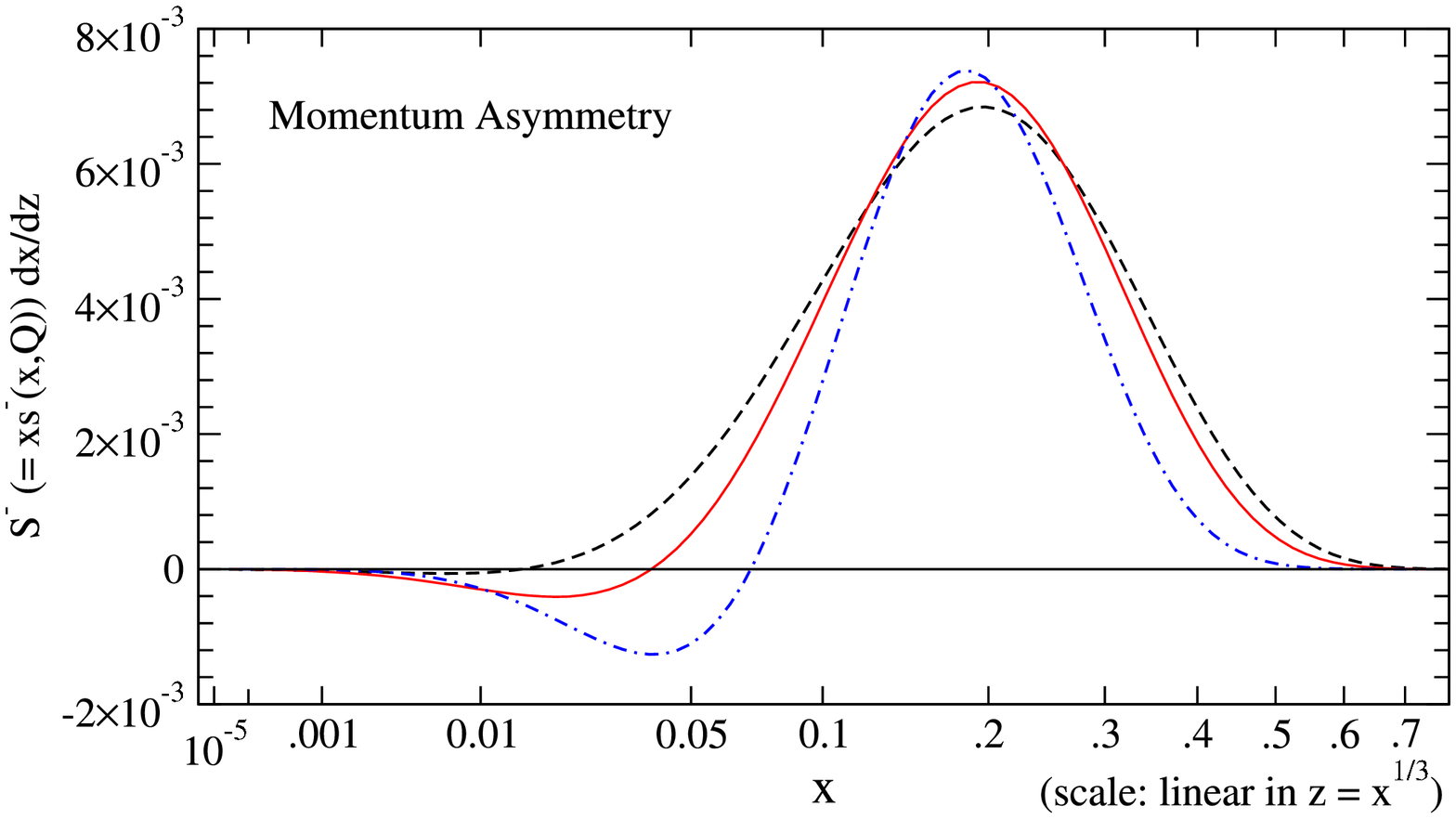}
\end{figure}
\begin{figure}
\hspace{0.4cm}
\includegraphics[width=.435\textwidth]{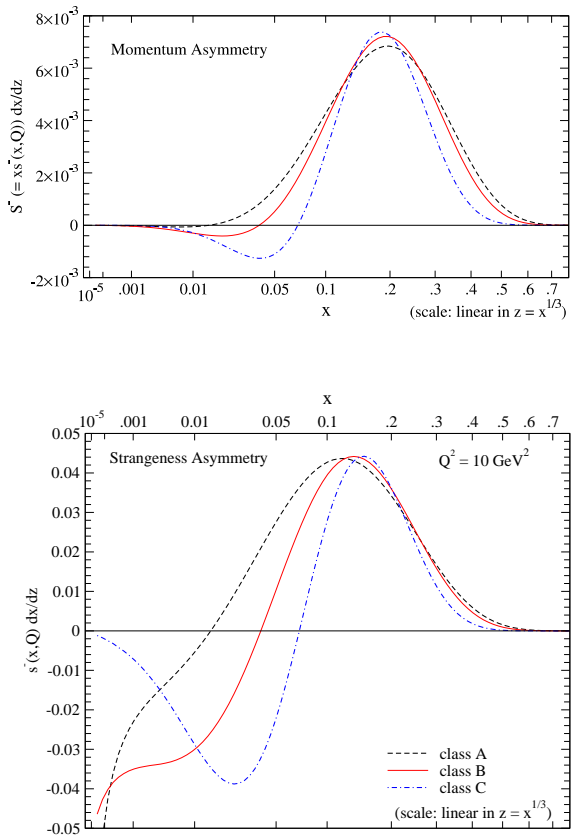}
\vspace{0.2cm}
\caption[]{CTEQ strange momentum asymmetry (top) and number asymmetry 
(bottom).}
\label{strasym}
\vspace{-0.5cm}
\end{figure}

MRST also look at the effect of {\bf isospin violation}\cite{cuts} since $R^-$ 
also depends 
on this -- 
$$
R^-=\frac{1}{2}-\sin^2 \theta_W +(1-\frac{7}{3}
\sin^2 \theta_W) \frac{[\delta U_{\rm v}] -[\delta D_{\rm v}]}{2[V^-]},
$$
where
$[\delta U_{\rm v}] = [U^p_{\rm v}] - [D^n_{\rm v}],$ 
$[\delta D_{\rm v}] = [D^p_{\rm v}] - [U^n_{\rm v}],$
and MRST use the simple parameterization
$$
u^p_{\rm v}(x) = d^n_{\rm v}(x) + \kappa f(x), \qquad 
d^p_{\rm v}(x) = u^n_{\rm v}(x) - \kappa f(x),
$$
where $f(x)$ is a simple function maintaining required conservation laws. The
dependence on $\kappa$ is shown in Fig.~6. The best fit value of 
$\kappa = -0.2$ leads to a similar reduction of the NuTeV anomaly, i.e. 
$\Delta sin^2\theta_W \sim -0.002$. But there is only a weak indication of 
this value and a fairly wide variation in $\kappa$ is allowed.

\begin{figure}[h!]
\begin{center}
\includegraphics[width=.5\textwidth]{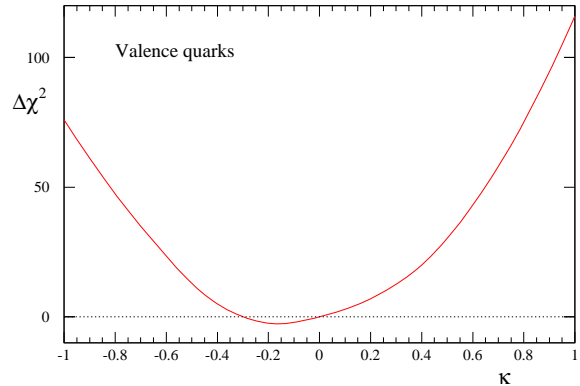}
\end{center}
\vspace{-0.4cm}
\caption[]{$\Delta \chi^2$ against the isospin violating parameter $\kappa$.}
\label{isospin}
\vspace{-0.5cm}
\end{figure}

The best determination of the {\bf high-$\bf x$ gluon
distribution} comes from inclusive
jet measurements by D0 and CDF at Tevatron.
They measure $d \sigma/dE_Td\eta$ for central rapidity CDF 
or in bins of rapidity D0. 
At central rapidity the kinematic equality (at LO) is $x= 2 E_T/\sqrt{s}$, 
and measurements extend up to
$E_T \sim 400 \GeV \,(x\sim 0.45)$, and down to 
$E_T \sim 60 \GeV \,(x \sim 0.06)$.  
Gluon-gluon fusion dominates the hard cross-section, 
but $g(x,\mu^2)$ falls off more quickly as $x \to 1$ than 
$q(x,\mu^2)$ so  there is 
a transition from gluon-gluon fusion at small $x$, to gluon-quark to 
quark-quark at high $x$. 
However, as seen in Fig.~7 even at the 
highest $x$ gluon-quark contributions are significant.
Jet photoproduction at HERA will be another
constraint in the future.

\begin{figure}[h!]
\begin{center}
\vspace{-0.6cm}
\hspace{-0.8cm}
\includegraphics[width=.46\textwidth]{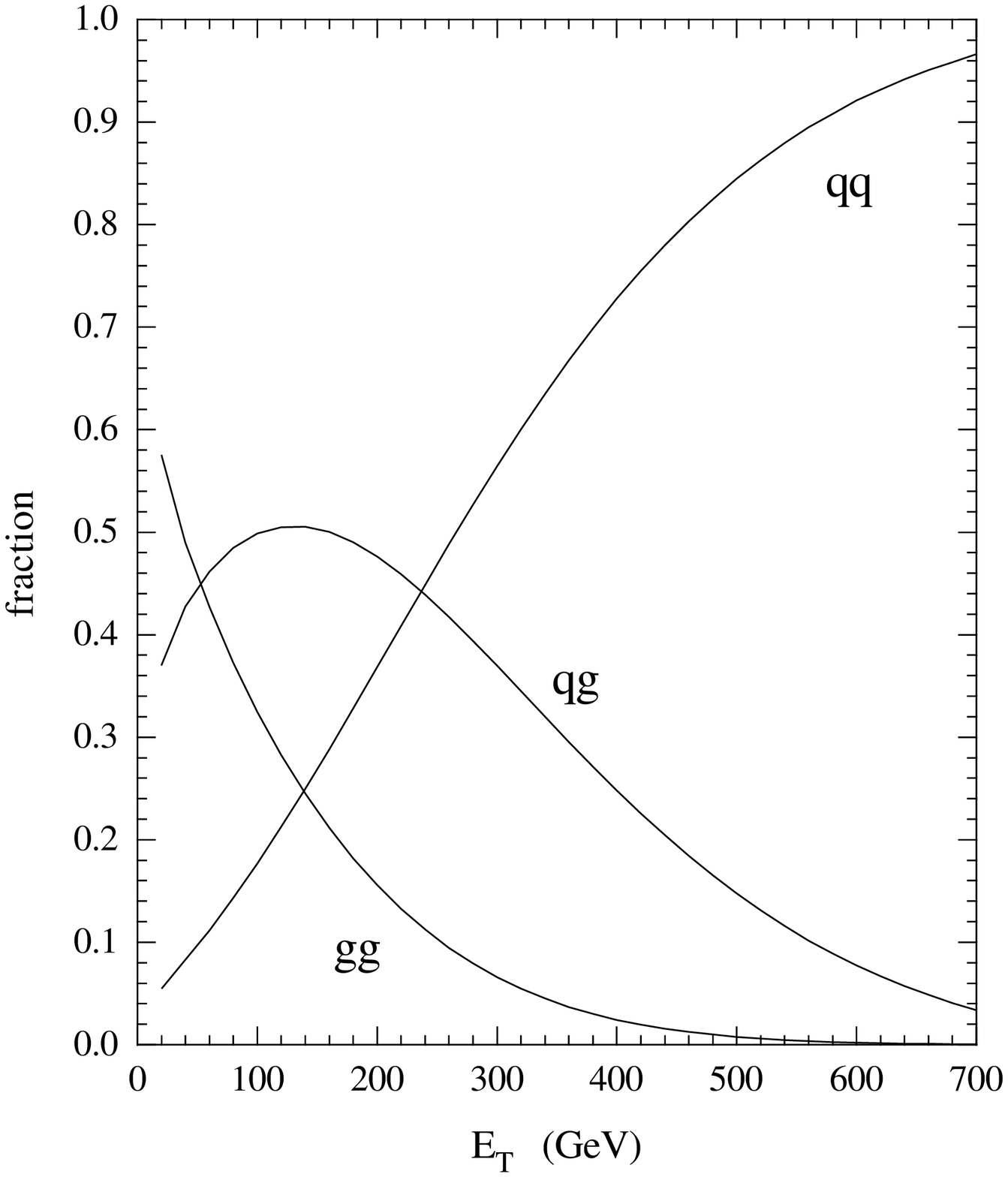}
\end{center}
\vspace{-2.6cm}
\caption[]{Fractional contributions to jet cross-section from different 
parton-level
contributions.}
\label{jetprops}
\vspace{-0.5cm}
\end{figure}

The above procedure completely 
determines the parton distributions at present.  
The total fit is reasonably good and that for CTEQ6\cite{CTEQ6} is shown in 
Table 1 for the large data sets. 
The total $\chi^2=1954/1811$. For MRST  
The total $\chi^2=2328/2097$
-- but the errors are treated differently, and different data sets and cuts
are used. The same sort of conclusion is true for other 
{\it global fits}\cite{Botje}$^-$\cite{ZEUSfit}  
(which use fewer data). 
However, there are some areas where the theory perhaps needs to be improved, 
as we will discuss later.

\begin{table}
\caption{Quality of fit to data for CTEQ6M.}
\begin{center}
\begin{tabular}{|ccc|}
\hline
Data Set & no. of data & $\chi^2$\\ 
\hline
H1 $ep$           & 230      & 228   \\                   
ZEUS $ep$         & 229      & 263    \\                   
BCDMS $\mu p$     & 339      & 378    \\
BCDMS $\mu d$     & 251      & 280    \\                   
NMC $\mu p$       & 201      & 305   \\                                
E605 (Drell-Yan) & 119      &  95   \\        
D0 Jets           &  90      &  65  \\ 
CDF Jets          &  33      &  49  \\
\hline
\end{tabular}
\end{center}
\label{Tab1a}
\vspace{-0.3cm}
\end{table}

\section{Parton Uncertainties} 

\subsection{Hessian (Error Matrix) approach} 

In this one defines the Hessian matrix $H$ by
$$ \chi^2 -\chi_{min}^2 \equiv \Delta \chi^2 = \sum_{i,j} 
H_{ij}(a_i -a_i^{(0)})
(a_j -a_j^{(0)}). $$
$H$ is related to the covariance matrix 
of the parameters by
$C_{ij}(a) = \Delta \chi^2 (H^{-1})_{ij},$
and one can use the standard formula for linear error propagation,  
$$(\Delta F)^2 = \Delta \chi^2 \sum_{i,j} \frac{\partial F}
{\partial a_i}(H)^{-1}_{ij}  
\frac{\partial F}{\partial a_j}.$$
This has been employed to 
find partons with errors by Alekhin\cite{Alekhin}, as seen in Fig.~8
and H1\cite{H1Krakow} (each with restricted data sets). 

\begin{figure}
\begin{center}
\includegraphics[width=.45\textwidth]{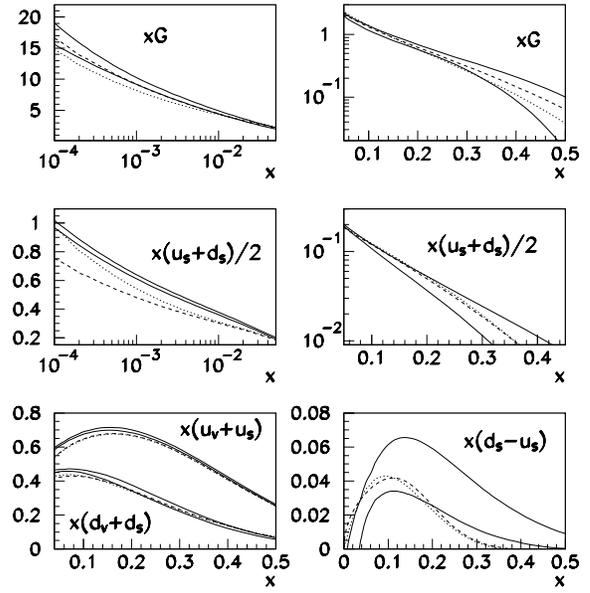}
\end{center}
\vspace{-0.1cm}
\caption[]{Results 
for Alekhin partons at $Q^2=9\,\GeV^2$ with uncertainties 
(solid lines), (dashed lines -- CTEQ5M, dotted lines -- MRST01).}
\vspace{-0.5cm}
\label{alekhin}
\end{figure}

The simple method can be problematic with larger data sets and 
larger numbers of parameters due to extreme 
variations in $\Delta \chi^2$ in different directions in parameter 
space. This is solved by finding and rescaling the eigenvectors of $H$ 
(CTEQ\cite{CTEQmul,CTEQHes,CTEQ6}) leading to the diagonal form 
$$\Delta \chi^2 = \sum_{i} z_i^2. $$
The uncertainty on a physical quantity is given by
$$(\Delta F)^2 = \sum_{i} \bigl(F(S_i^{(+)})-F(S_i^{(-)})\bigr)^2,$$
where $S_i^{(+)}$ and $S_i^{(-)}$ are PDF sets 
displaced along eigenvector
directions by a given $\Delta \chi^2$. Similar eigenvector 
parton sets have also been introduced by MRST\cite{MRSTnew} and ZEUS.  
However, there is an art in choosing the ``correct'' 
$\Delta \chi^2$ given the complication of the errors in the full 
fit\cite{THJPG}. 
Ideally $\Delta \chi^2 = 1$, but this leads to unrealistic errors,
e.g. values of $\alpha_S(M_Z^2)$ obtained by CTEQ
using $\Delta \chi^2 =1$ for each data set in the global fit
are shown in Fig.~9, and are not consistent. 
CTEQ choose $\Delta \chi^2 
\sim 100$, which is perhaps conservative. 
MRST choose $\Delta \chi^2 \sim 50$. An example of results 
is shown in Fig.~10.

\begin{figure}[h!]
\vspace{2.4cm}
\hspace{-1.2cm}
\includegraphics[width=.4\textwidth]{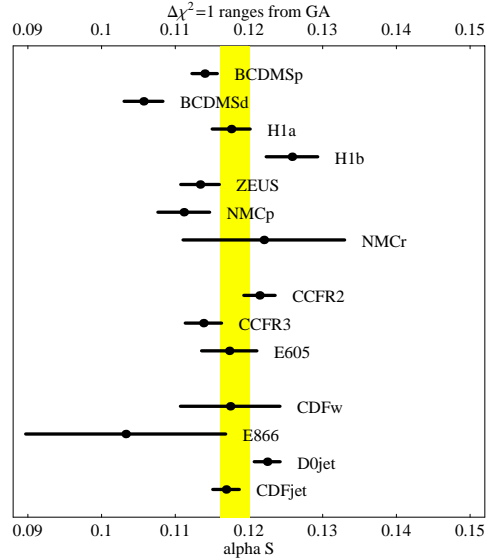}
\vspace{-2.8cm}
\caption[]{Values of $\alpha_S(M_Z^2)$ and their uncertainties 
using $\Delta\chi^2=1$ from CTEQ.}
\label{alphachicteq}
\vspace{-0.3cm}
\end{figure}

\begin{figure}[h!]
\vspace{-2.2cm}
\includegraphics[width=.4\textwidth]{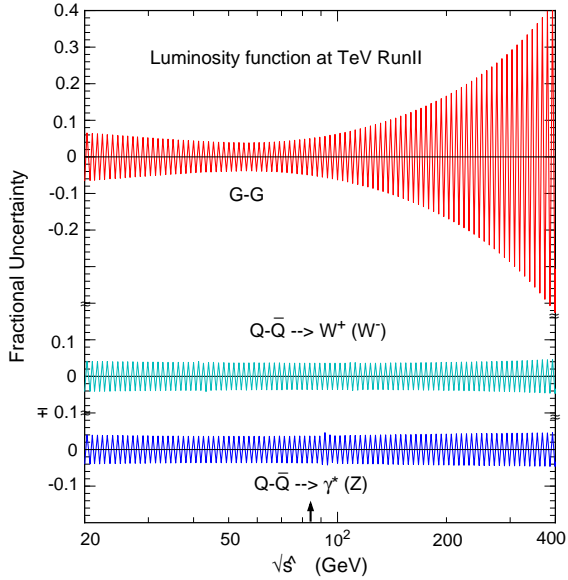}
\vspace{-0.4cm}
\caption[]{Luminosity 
uncertainties using the CTEQ Hessian approach.}
\label{CTEQlum}
\vspace{-0.8cm}
\end{figure}

\subsection{Offset method}  

In this the best fit and parameters 
$a_0$ are obtained using only uncorrelated 
errors. The quality of the fit is then 
estimated by adding uncorrelated and correlated errors in quadrature. 
Roughly speaking systematic uncertainties are 
determined by letting each source of systematic error vary by
$1\sigma$ and adding the deviations in 
quadrature. This procedure is used by ZEUS\cite{ZEUSfit}, 
and leads to  an effective $\Delta \chi^2 >1$. 

\begin{figure}[h!]
\begin{center}
\vspace{-0.3cm}
\includegraphics[width=.5\textwidth]{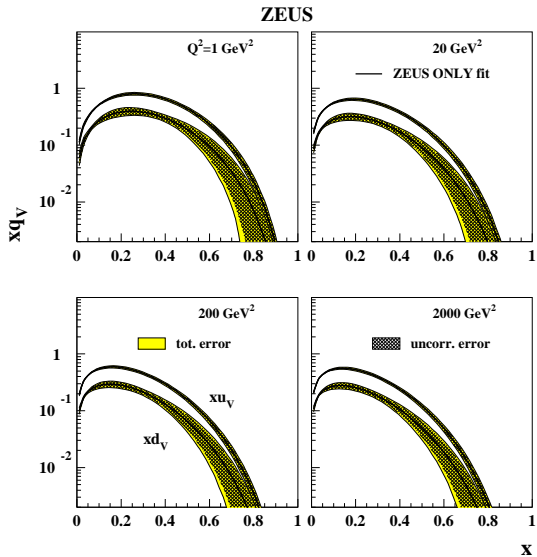}
\end{center}
\vspace{-0.8cm}
\caption[]{The valence partons extracted by ZEUS from a {\it global} 
fit and a fit to 
their own data alone (with some input assumptions). The latter illustrates a 
potential for a real constraint from HERA data alone in the future.}
\label{offset}
\vspace{-0.3cm}
\end{figure}

\subsection{Statistical Approach} 

In principle this involves the 
construction of an ensemble of distributions labelled by ${\cal F}$ each with 
probability $P(\{{\cal F}\})$, where one can incorporate the full 
information about measurements and their error 
correlations into the calculation of $P(\{{\cal F}\})$. 
This is statistically correct, and does not rely on 
the approximation of linear propagation errors in calculating observables.
However, it is inefficient, and in practice 
one generates $N$ ($N$ can be as low as $100$) 
different distributions
with unit weight but distributed according to $P(\{{\cal F}\})$\cite{Giele}. 
Then the mean $\mu_O$ and deviation $\sigma_O$ of an observable $O$ are  
given by
$$\mu_O = \frac {1}{N}\sum_1^{N}O(\{{\cal F}\}),\,
\sigma_O^2 =\frac {1}{N} \sum_{1}^{N}(O(\{{\cal F}\})-\mu_O)^2.
$$

Currently this approach uses only proton DIS data sets in order to avoid 
complicated 
uncertainty issues, e.g. shadowing effects for nuclear targets, 
and also demands consistency between data sets.
However, it is difficult to find 
many truly compatible DIS experiments, and consequently the 
Fermi2001 partons are determined by only 
H1, BCDMS, and E665 data sets. 
They result in good predictions for many Tevatron cross-sections,
e.g. inclusive jets and $W$ and $Z$ total cross-sections. 
However, the restricted data sets mean there is restricted information --
data sets are deemed either perfect or, in the case of  most of them,  
useless -- leading to unusual values for some parameters. e.g.  
$\alpha_S(M_Z^2)=0.112 \pm 0.001$ and a very hard $d_V(x)$ 
at high $x$ (together these two features facilitate 
a good fit to Tevatron jets 
independent of the high-$x$ gluon). 
These partons would produce some extreme predictions, as seen later.
Nevertheless, the approach does demonstrate that the Gaussian approximation 
is often not good, and therefore highlights shortcomings in the methods 
outlined in the previous sections.
It is a very attractive, but ambitious large-scale project, still in need 
of some further development. In particular I feel it requires the inclusion of 
a wider variety of data in order to overcome the obstacle presented by the 
fact that most data sets in the global fit are not really as consistent
as they should be in the strict statistical sense.     

\subsection{Lagrange Multiplier method} 

This was first suggested by CTEQ\cite{CTEQLag} and 
has been concentrated on by MRST\cite{MRSTnew}. One performs the 
fit while constraining the value of some physical quantity, i.e. one minimizes 
$$ \Psi(\lambda,a) = \chi^2_{global}(a)  + \lambda F(a)$$
for various values of $\lambda$. This gives a set of best fits for 
particular 
values of the quantity $F(a)$ without relying on the quadratic approximation 
for $\chi^2$, as shown for $\sigma_W$ in Fig.~12. 
The uncertainty is then determined by deciding an allowed range of 
$\Delta \chi^2$.
One can also easily check the variation in $\chi^2$ 
for each of the experiments in the global fit and ascertain if the total
$\Delta\chi^2$ is coming specifically from one region, which might cause 
concern. In principle, this is superior to the Hessian approach, but it must 
be repeated for each physical process. 

\begin{figure}[h!]
\hspace{-0.3cm}
\includegraphics[width=.42\textwidth]{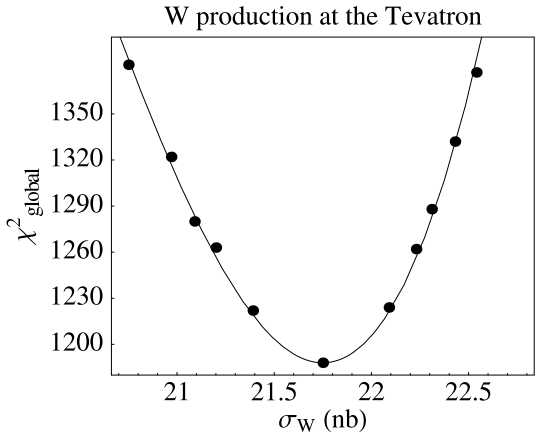}
\vspace{-0.1cm}
\caption[]{$\chi^2_{global}$ for CTEQ plotted against $\sigma_W$.}
\label{lagrange}
\vspace{-0.7cm}
\end{figure}

\subsection{Results}

I choose the cross-section for $W$ and Higgs production at the Tevatron and 
LHC (for $M_H=115\GeV$) as examples.  
Using their fixed value of $\alpha_S(M_Z^2) =0.118$ and $\Delta \chi^2=100$ 
CTEQ obtain 
$$\Delta \sigma_{W}(\rm LHC) \approx \pm 4\% \quad 
\Delta \sigma_{W}({\rm Tev}) \approx \pm 5\%$$
\vspace{-0.3cm}
$$ \Delta 
\sigma_{H}({\rm LHC}) \approx \pm 5\%.$$
Using a slightly wider range of data, 
$\Delta \chi^2 \sim 50$ and $\alpha_S(M_Z^2) =0.119$ MRST obtain
$$\Delta \sigma_{W}(\rm Tev) \approx \pm 1.2\% \quad 
\Delta \sigma_{W}({\rm LHC}) \approx \pm 2\%$$
\vspace{-0.3cm}
$$\Delta \sigma_{H}(\rm Tev) \approx \pm 4\% \quad 
\Delta \sigma_{H}({\rm LHC}) \approx \pm 2\%.$$
MRST also allow $\alpha_S(M_Z^2)$ to be free. 
In this case $\Delta \sigma_{W}$ is quite 
stable but $\Delta \sigma_{H}$ almost doubles. Contours of variation in 
$\chi^2$ for the predictions of these cross-sections are shown in Fig.~13.    

\begin{figure}[h!]
\includegraphics[width=.38\textwidth]{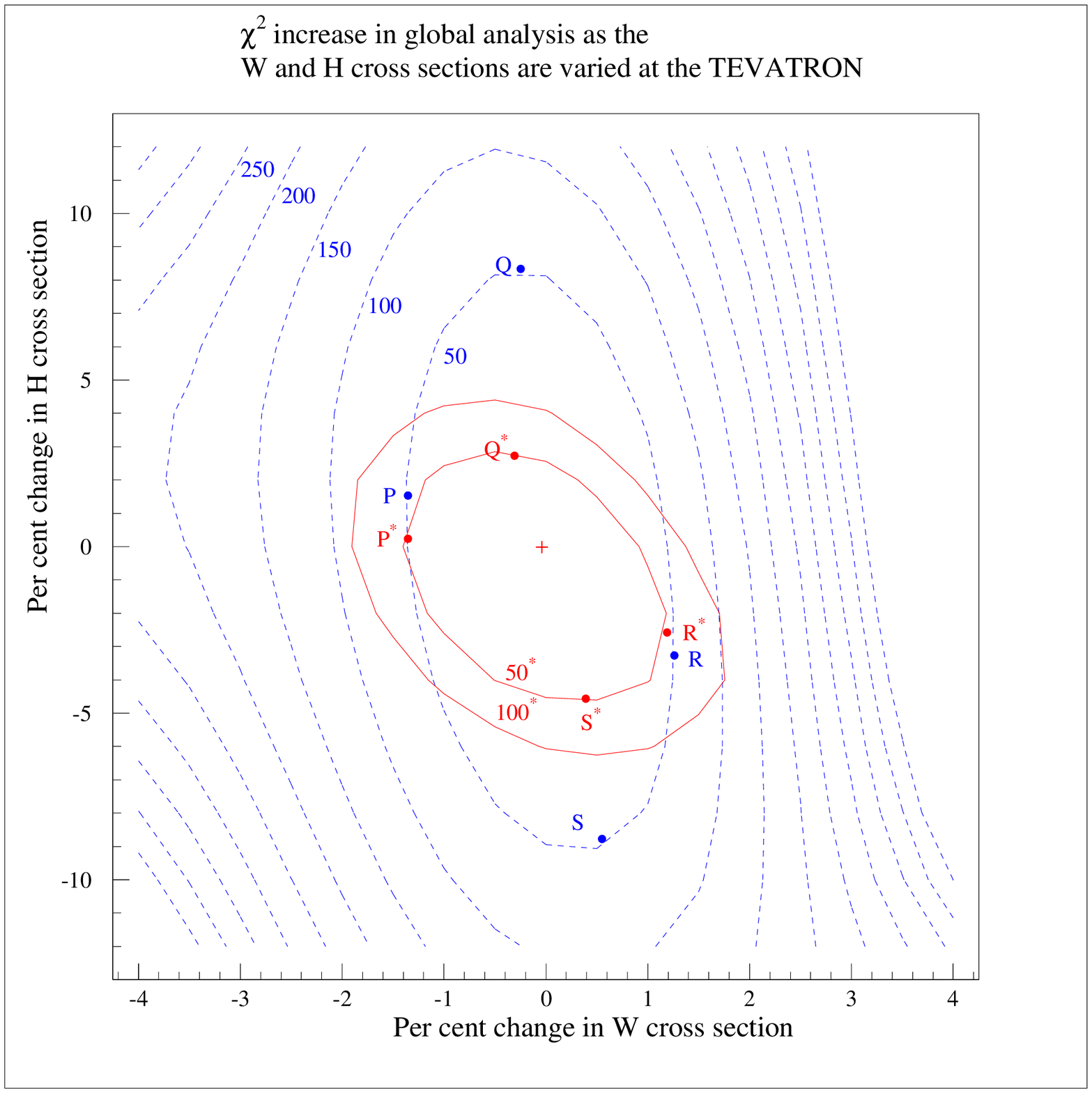}
\vspace{-1.8cm}
\end{figure}
\begin{figure}[h!]
\includegraphics[width=.38\textwidth]{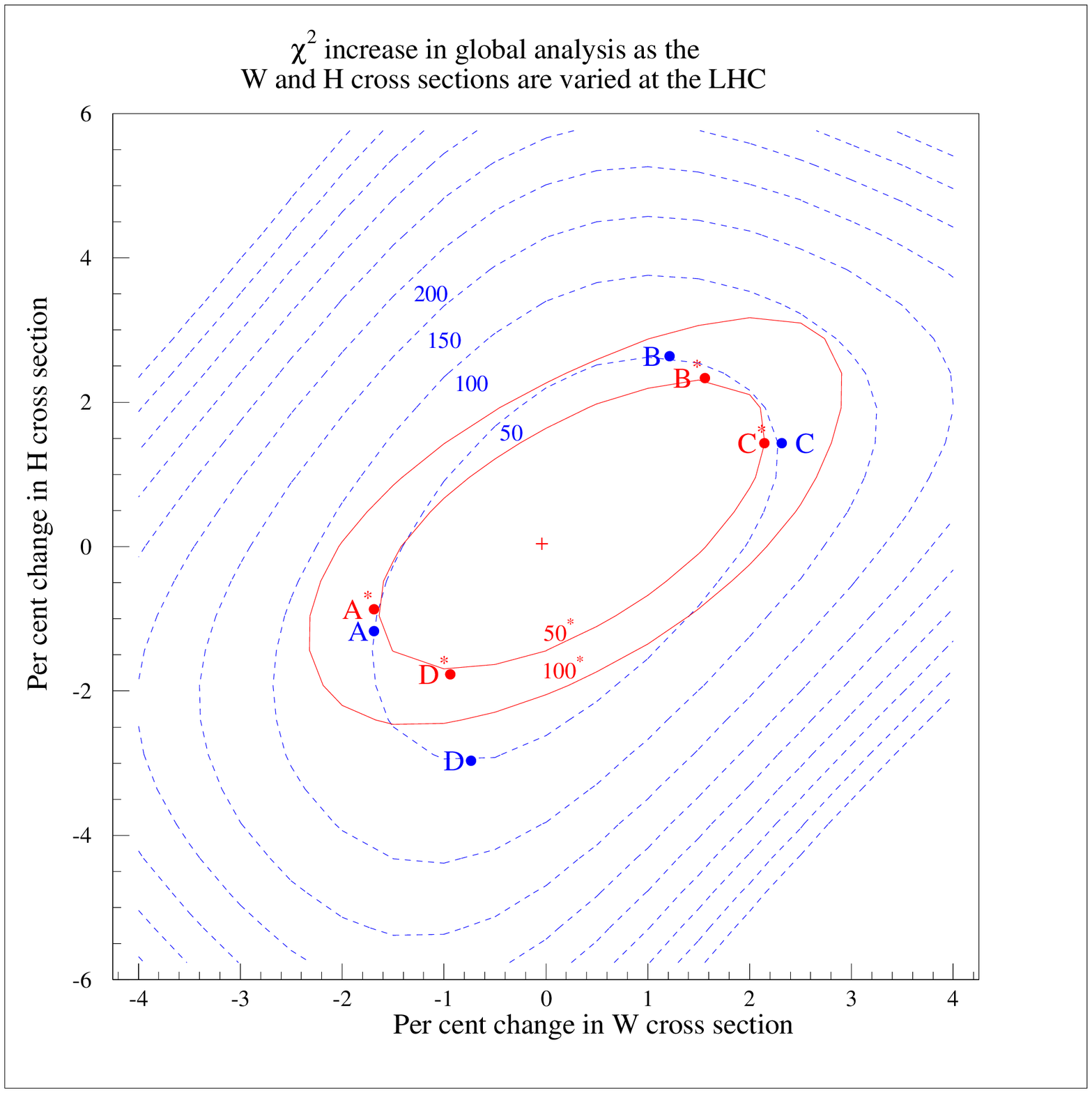}
\vspace{-1.5cm}
\caption[]{$\chi^2$-plot for $W$ and Higgs production at the 
Tevatron (top) and LHC (bottom) 
with $\alpha_S$ free (dashed) and fixed (solid) at $\alpha_S=0.119$}
\label{eps15}
\vspace{-0.5cm}
\end{figure}

The same general procedure is also used by CTEQ\cite{CTEQnew} 
to look at the effect
of new physics parameterized by the contact term
$$ \pm(2\pi/\Lambda^2)
(\bar q_L\gamma^{\mu}q_L)(\bar q_L\gamma_{\mu}q_L).$$
The curves in Fig.~14 show the fit to the $D0$ jet data, which is the most
discriminating data set, for $\Lambda =1.6,
2.0, 2.4, \infty \,\,\hbox{\rm TeV}$, and $A=-1$.   
For the highest values of $\Lambda$ the fit even improves very slightly, but 
$\Lambda > 1.6,\,\hbox{\rm TeV}$ is clearly ruled out.

\begin{figure}[h!]
\includegraphics[width=.43\textwidth]{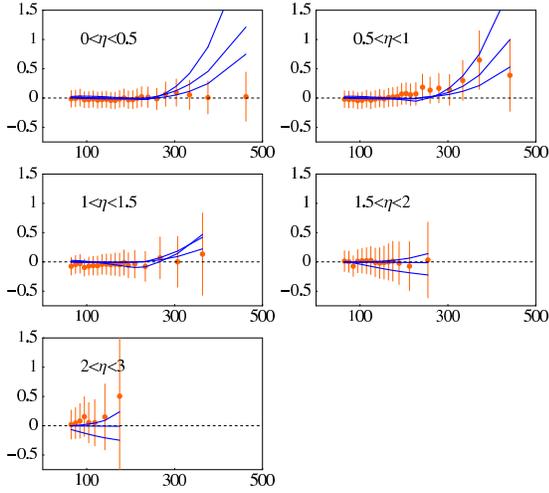}
\vspace{-0.0cm}
\caption[]{Fits to D0 jets for different values of $A$.}
\label{newphys}
\vspace{-0.8cm}
\end{figure}

\begin{table*}
\caption{Values of $\alpha_s(M_Z^2)$ and its error from different NLO QCD 
fits.}
\begin{center}
{\begin{tabular}{|l|c|ll|}
\hline
Group & $\Delta \chi^2$ & $\alpha_S(M_Z^2)$ &\\
\hline
  CTEQ6 & $\Delta \chi^2 = 100$ & $\alpha_s(M_Z^2)=$
    &$0.1165\pm 0.0065(exp)$\hfill\\ 
  ZEUS  & $\Delta \chi_{eff}^2 = 50$  
  &$\alpha_s(M_Z^2)=$&$ 0.1166
  \pm 0.0049(exp) $ 
  $\pm 0.0018(model)$ 
  $\pm 0.004(theory) $\\ 
  MRST01 & $\Delta \chi^2 = 20$ & $\alpha_s(M_Z^2)=$
  &$ 0.1190
  \pm 0.002(exp)$ 
  $\pm 0.003(theory)  $ \\
  H1    & $\Delta \chi^2 = 1$ & $\alpha_s(M_Z^2)=$& 
  $0.115 \pm 0.0017(exp)$ $ ^{+~~0.0009}_{-~~0.0005}~(model)$ 
  $\pm 0.005(theory)$ \\
  Alekhin & $\Delta \chi^2 = 1$ & $\alpha_s(M_Z^2)=$&
   $0.1171 
  \pm 0.0015(exp)$ $ \pm 0.0033(theory)$ \\
 GKK & CL & $\alpha_s(M_Z^2)=$& $0.112 
  \pm 0.001(exp)$ \\

\hline
\end{tabular} 
}
\label{tab:alf}
\end{center}
\vspace{-0.5cm}
\end{table*}

Hence, the estimation of uncertainties due to experimental errors 
has many different approaches and different types and amount of data actually 
fit. Overall the uncertainty from this source 
is rather small --  
only more than a few $\%$ for quantities determined by  
the high $x$ gluon and very high $x$ down quark. This is 
illustrated for the determinations of $\alpha_S(M_Z^2)$ in Table 2. There is 
generally good agreement, but their are some outlying values.

These outlying values of $\alpha_S(M_Z^2)$ show that
different approaches can sometimes lead to rather different central values,
This suggests that there are other matters to consider   
as well as the experimental errors on data. We also need to determine 
the effect of assumptions made about the fit, e.g. 
cuts made on the data, the data sets fit, the parameterization for input sets, 
the form of the strange sea, {\it etc}.. 
Many of these can be as important as the 
errors on the data used (or more so). This is demonstrated by 
the results from the LHC/LP Study Working Group\cite{Bourilkov} 
shown in Tables 3, and by predictions for $\sigma_W$ by MRST
CTEQ and Alekhin\cite{alekhinw} in Table 4. 
In both cases the discrepancies are mainly 
due to differences in detailed constraints (by data) on the quark 
decomposition. Differences between predictions are also shown by
Fig.~15 -- the predictions for 
$W$ and Higgs production at the Tevatron from MRST2001 and CTEQ6, and Fig.~16
-- the comparison between the gluons for the two parton sets.

\begin{table}
\caption{Cross sections for Drell-Yan pairs ($e^+e^-$) with 
{\tt PYTHIA} 6.206,
rapidity $<2.5$.
The errors shown are the PDF uncertainties.}
\vspace{-0.4cm}
\label{tab:zpole}
  \begin{center}
\begin{tabular}{|l|l|r|c|}
\hline
PDF set   &  Comment                    &  xsec [pb] & PDF uncertainty \%    \\
\hline
             \multicolumn{4}{|c|}{$81 < M < 101$ GeV}    \\
\hline
CTEQ6     &  LHAPDF                     &  1065 $\pm$ 46  &  4.4 \\
MRST2001  &  LHAPDF                     &  1091 $\pm$ ... &  3    \\
Fermi2002 &  LHAPDF   &  853 $\pm$ 18  & 2.2 \\
\hline
    \end{tabular}
  \end{center}
\vspace{-0.3cm}
\end{table}

\begin{table}
\caption{Comparison of $\sigma_W\cdot B_{l\nu}$ 
for different partons.}   
\label{tab:sigw}
  \begin{center}
\begin{tabular}{|l|l|r|c|}
\hline
PDF set   &  Comment                    &  xsec [nb] & PDF uncertainty    \\
\hline
Alekhin   &  Tevatron    & 2.73 & $\pm$ 0.05 (tot) \\
MRST2002  &  Tevatron           &  2.59 & $\pm$ 0.03 (expt)\\
CTEQ6  &  Tevatron           &  2.54 & $\pm$ 0.10 (expt)\\
\hline
Alekhin &  LHC            &  215 & $\pm$ 6 (tot) \\
MRST2002  &  LHC                   &  204 & $\pm$ 4 (expt) \\
CTEQ6  &  LHC                   &  205 &  $\pm$ 8 (expt) \\
\hline
    \end{tabular}
  \end{center}
\vspace{-0.3cm}
\end{table}

\begin{figure}[h!]
\includegraphics[width=.38\textwidth]{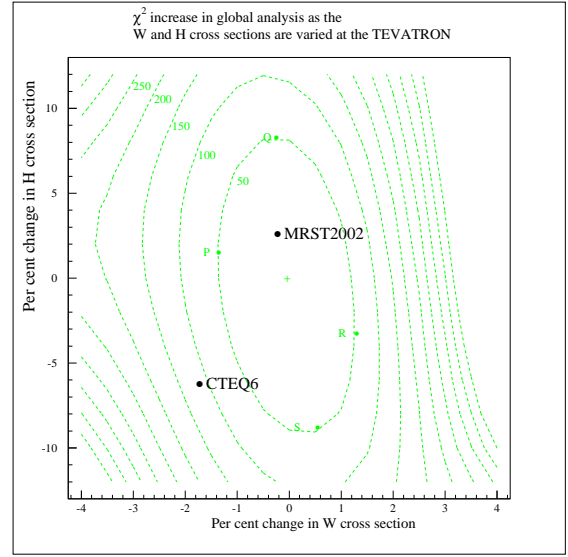}
\vspace{-1.8cm}
\caption[]{$\chi^2$-plot for $W$ and Higgs production at the Tevatron
with $\alpha_S$ free. The predictions from CTEQ6 is marked.}
\label{cteqvmrstwh}
\end{figure}

\begin{figure}[h!]
\hspace{-1cm}
\includegraphics[width=.45\textwidth]{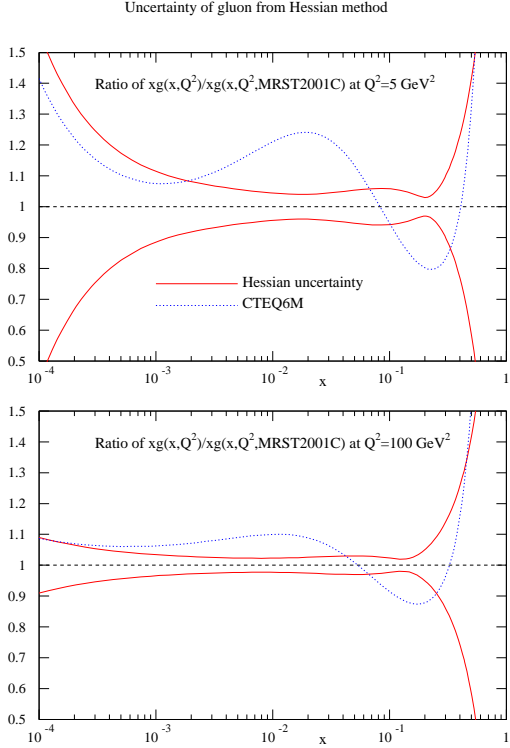}
\vspace{-1.7cm}
\caption[]{Fractional uncertainty in the MRST gluon compared with the 
difference in the central CTEQ6 gluon.}
\label{cteqvmrstg}
\vspace{-1.0cm}
\end{figure}

\vspace{-0.2cm}

\section{Theoretical errors}

\vspace{-0.2cm}

\subsection{Problems in the fit}

As well as 
the consequences of these assumptions we must consider 
the related problem of theoretical errors.
Theoretical errors are indicated by some regions where the theory perhaps 
needs to be improved to fit the data better.
There is a reasonably good fit to HERA data, 
but there are some problems at the highest 
$Q^2$ at moderate $x$, i.e. in $d F_2/d \ln Q^2$, as seen for MRST and CTEQ 
in Fig.~17. Also the data require the gluon to be valencelike or negative at 
small $x$ at low $Q^2$, e.g. the ZEUS gluon in Fig.~18, leading to 
$F_L(x,Q^2)$ being negative\cite{MRST2001} at the smallest $x,Q^2$.
However, it is not just the low $x$--low $Q^2$ data that require 
this negative gluon. 
The moderate $x$ data need lots of gluon to get a reasonable 
$d F_2/d \ln Q^2$ and the Tevatron jets need a large high $x$ gluon, 
and this must be compensated for elsewhere.  
In general MRST find that it is difficult to reconcile the fit to 
jets and to the rest of the data, Fig.~19, and that   
different data compete over the gluon and $\alpha_S(M_Z^2)$.
The jet fit is better for CTEQ6 largely due to their 
different cuts on other data. Other fits do not include the Tevatron jets,
but generally produce gluons largely incompatible with this data.   

\begin{figure}
\begin{center}
\vspace{-0.8cm}
\includegraphics[width=.48\textwidth]{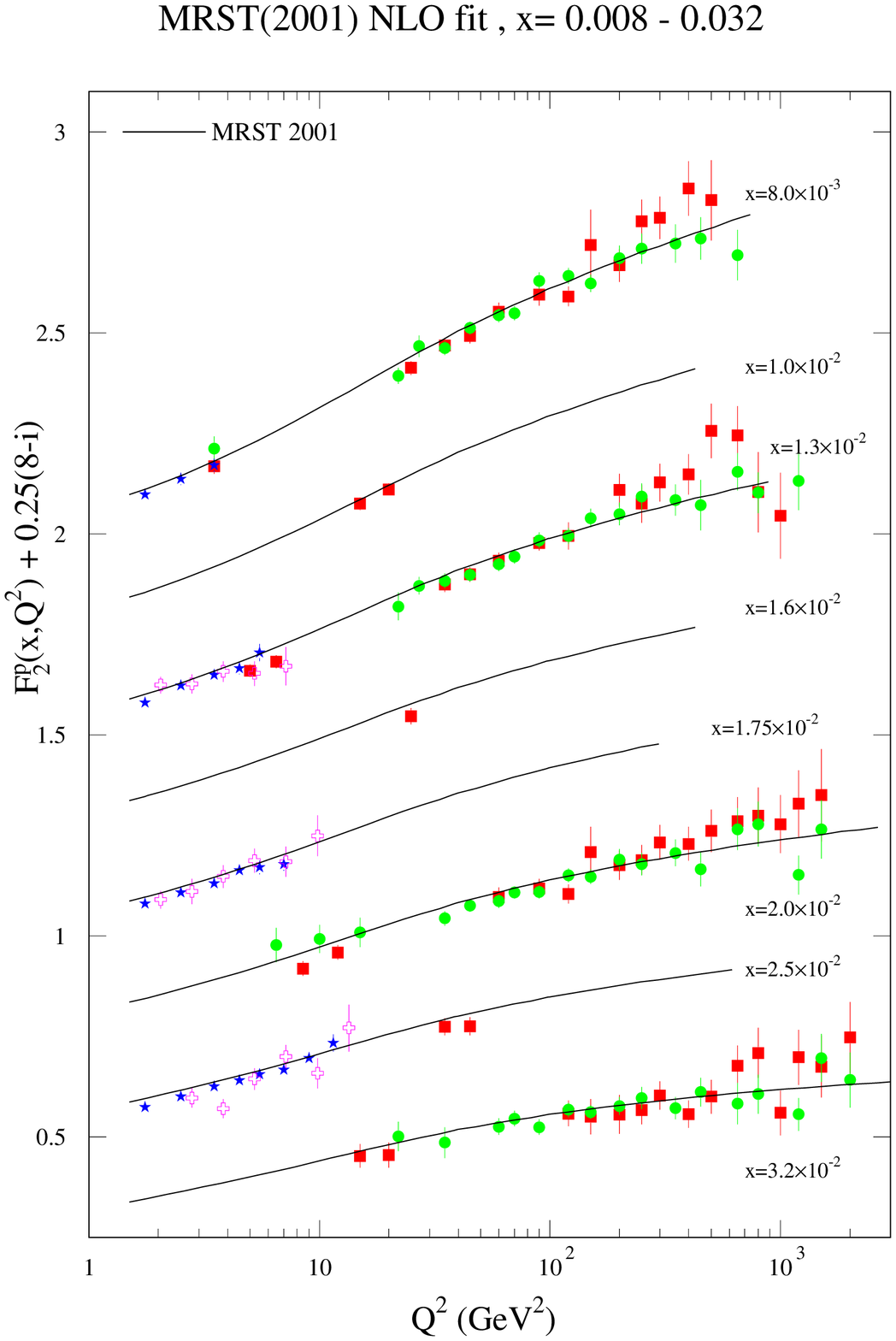}
\end{center}\end{figure}
\begin{figure}
\begin{center}
\vspace{-0.3cm}
\includegraphics[width=.48\textwidth]{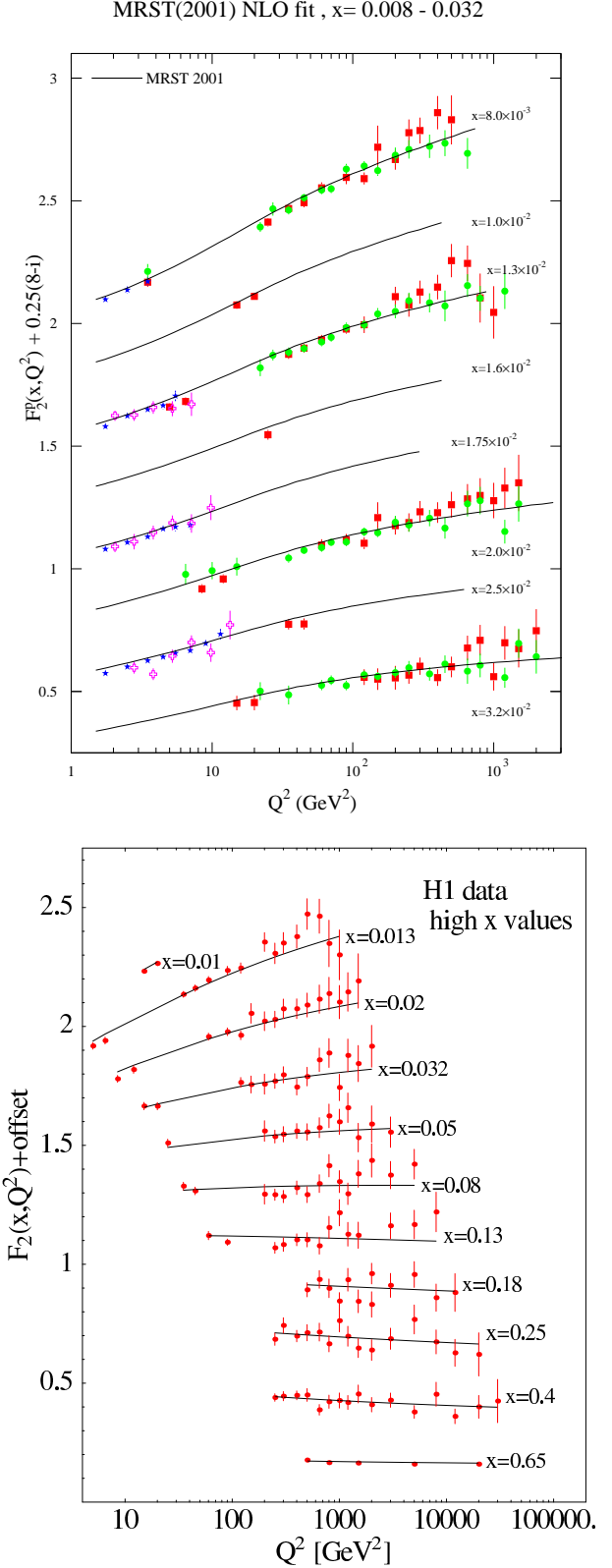}
\end{center}
\vspace{-0.3cm}
\caption[]{Comparison of MRST(2001) $F_2(x,Q^2)$ with HERA,
NMC and E665 data (top) and  CTEQ6 $F_2(x,Q^2)$ with H1 data
(bottom).}
\vspace{-0.3cm}
\label{eps21}
\end{figure}

\begin{figure}[h!]
\begin{center}
\includegraphics[width=.47\textwidth]{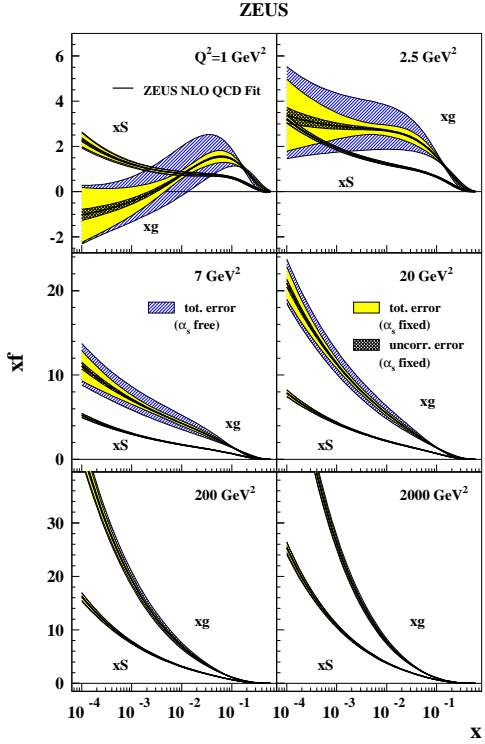}
\end{center}
\vspace{-0.72cm}
\caption[]{Zeus gluon and sea quark distributions at various $Q^2$ values.}
\label{eps24}
\vspace{-0.0cm}
\end{figure}

\begin{figure}[h!]
\hspace{-0.8cm}
\includegraphics[width=.45\textwidth]{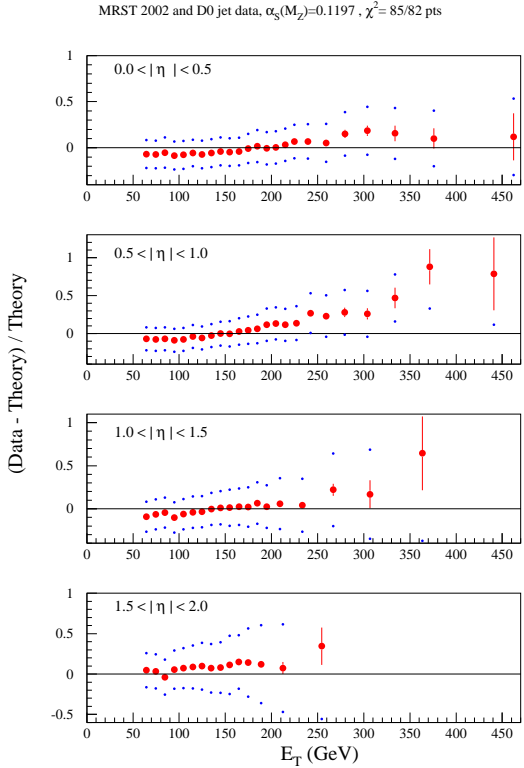}
\vspace{-1.4cm}
\caption[]{The MRST fit to D0 jet data. The points show the range of 
the systematic errors.}
\label{d0}
\vspace{-0.4cm}
\end{figure}

\subsection{Types of Theoretical Error, NNLO}

It is vital to consider theoretical errors. These include 
higher perturbative orders (NNLO), small $x$ ($\alpha_s^n \ln^{n-1}(1/x)$),
large $x$ ($\alpha_s^n \ln^{2n-1}(1-x)$)
low $Q^2$ (higher twist), {\it etc}..
Note that renormalization/factorization 
scale variation is not a reliable  method of estimating these 
theoretical errors because of increasing 
logs at higher orders.  

\begin{figure}[h!]
\begin{center}
\vspace{0.1cm}
\includegraphics[width=.5\textwidth]{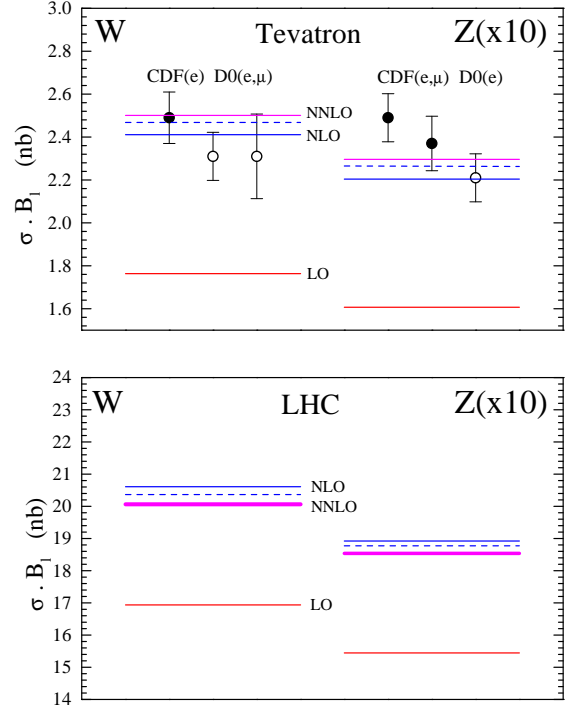}
\end{center}
\vspace{-1.7cm}
\caption[]{LO, NLO and NNLO predictions for $W$ and $Z$ cross-sections.}
\vspace{-0.5cm}
\label{eps29}
\end{figure}

In order to investigate the true theoretical error we 
must consider some way of 
performing correct large and small $x$ resummations,
and/or use what we already know about NNLO. 
The coefficient functions are known at NNLO.
Singular limits $x \to 1$, $x \to 0$ are known for NNLO
splitting functions as well as limited moments\cite{NNLOmoms}, and this
has allowed approximate NNLO 
splitting functions to be devised\cite{NNLOsplit}
which have been used in approximate global fits\cite{MRSTNNLO}.
They improve the quality of fit very slightly  
(mainly at high $x$) and 
$\alpha_S(M_Z^2)$ lowers from $0.119$ to 0.1155. 
The gluon is smaller at NNLO at low $x$ due to the positive NNLO 
quark-gluon splitting function. 
There is also a NNLO fit by Alekhin\cite{NNLOAl}, with some differences 
-- the gluon is not smaller, probably due 
to the absence of Tevatron jet data in the fit and to a very
different definition of the NNLO charm contribution. 
There is agreement in the reduction of $\alpha_S(M_Z^2)$ at NNLO, i.e. 
$0.1171 \to 0.1143$.

Using these NNLO partons there is reasonable stability order by order for 
the (quark-dominated) $W$ and $Z$ cross-sections, as seen in Fig.~20. 
However, the change from NLO to NNLO is of order $4\%$, which is much 
bigger than the uncertainty at NLO due to experimental errors. 
Also, this fairly good convergence is largely 
guaranteed because the quarks are fit directly to data.
There is greater danger in gluon dominated quantities, e.g. $F_L(x,Q^2)$, as 
can be seen in Fig.~21. Hence, the convergence from 
order to order is uncertain.

\begin{figure}
\vspace{-0.5cm}
\hspace{-1cm}
\includegraphics[width=.45\textwidth]{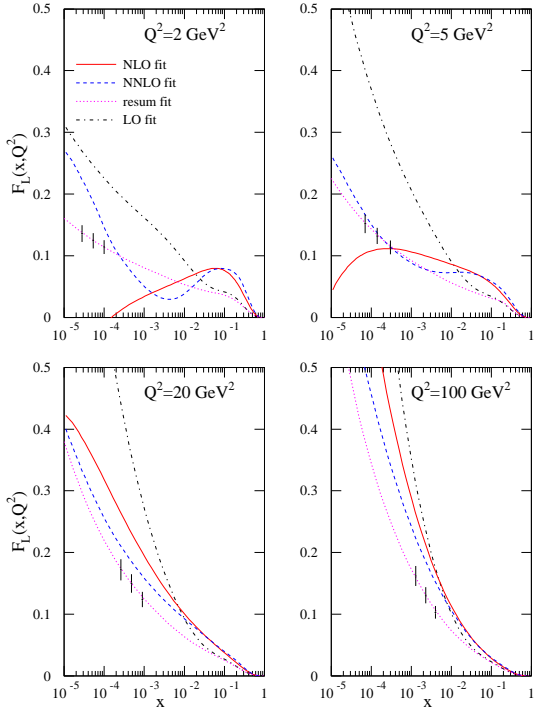}
\vspace{-1.6cm}
\caption[]{Comparison of the predictions for $F_L(x,Q^2)$ at LO, 
NLO and NNLO using MRST partons and also a  
$\ln(1/x)$-resummed prediction\cite{RTsum}.}
\vspace{-0.6cm}
\label{hera3}
\end{figure}

\vspace{-0.3cm}

\subsection{Empirical approach} 

We can estimate where theoretical errors may be important by adopting 
the empirical 
approach of investigating in detail the effect of cuts on the 
fit quality, i.e. we try varying the kinematic cuts on data. 
The procedure is to change $W^2_{cut}$, $Q^2_{cut}$ 
and/or $x_{cut}$, 
re-fit and see if the quality of the fit to the remaining data improves 
and/or the input parameters change dramatically. 
(This is similar to a previous suggestion in terms of 
data sets\cite{Collins}.) 
One then continues until the quality of the fit and the partons 
stabilize\cite{cuts}.

For $W^2_{cut}$ raising from $15 \GeV^2$ has no effect.
When raising $Q^2_{cut}$ from $2\GeV^2$ in steps there 
is a  slow, continuous and significant improvement for $Q^2$ up to 
$> 10 \GeV^2$ (560 data points cut), suggesting that 
any corrections are probably higher orders not higher twist. 
The input gluon becomes slightly smaller at low $x$
at each step (where one loses some of the lowest $x$ data), 
and larger at high $x$. 
$\alpha_S(M_Z^2)$ slowly decreases by about 0.0015.
Raising $x_{cut}$ leads to continuous improvement with  
stability reached at $x=0.005$ (271 data points cut) with 
$\alpha_S(M_Z^2) \to 0.118$. There is  
an improvement in the fit to HERA, NMC and Tevatron jet data, and 
much reduced tension between the data sets.
At each step the moderate $x$ gluon becomes more 
positive, at the expense of the gluon below the cut becoming very negative 
and $dF_2(x,Q^2)/d\ln Q^2$ being incorrect. However, higher orders could cure 
this in a quite plausible manner. For example adding higher order terms to 
the splitting functions 
$$ P_{gg} \to ....+ \frac{3.86\bar\alpha_S^4}{x}
\biggr(\frac{\ln^3(1/x)}{6}-
\frac{\ln^2(1/x)}{2}\biggr),$$
\vspace{-0.5cm}
$$ P_{qg} \to ....+ \frac{5.12N_f\bar\alpha_S^5}{6x}
\biggr(\frac{\ln^3(1/x)}{6}-
\frac{\ln^2(1/x)}{2}\biggr),$$
leaves the improved 
fit above $x=0.005$ largely unchanged, but solves the problem 
below $x=0.005$. Saturation corrections added to NLO and NNLO fits 
seem to make the situation worse.   
Hence, the cuts are suggestive of theoretical errors for 
small $x$ and/or small $Q^2$. 
Predictions for $W$ and Higgs cross-sections at the Tevatron are still safe if 
$x_{cut}=0.005$, since they do not sample partons at lower $x$. However, 
they change in a smooth manner as $x_{cut}$ is lowered, due to the 
altered partons above $x_{cut}$.
 
There is a lot of work on explicit $\ln(1/x)$-resummations in structure 
functions 
and parton distributions for example\cite{RTsum,CCSS,ABF},
but there is no complete consensus on the best approach. 
There is also work on connecting the partons to alternative 
approaches at small $x$, e.g.
dipole models\cite{dipole}, and pomerons\cite{DL}.  
These approaches can suggest improvements to the fits and changes 
in predictions, e.g. a resummed prediction\cite{RTsum} for 
$F_L(x,Q^2)$ is shown on Fig.~21.
Accurate and direct measurements of $F_L(x,Q^2)$ and other 
quantities at low $x$ 
and/or $Q^2$ (the predicted range and accuracy of $F_L(x,Q^2)$
measurements at HERA III is shown on Fig.~21) 
would be a great help in determining whether
$NNLO$ is sufficient or whether resummed  (or other) corrections are
necessary, or helpful for maximum precision.

\vspace{-0.4cm}

\section{Conclusions}

\vspace{-0.2cm}

One can perform global fits to all up-to-date data over a wide range of 
parameter space, and  
there are various ways of looking at uncertainties
due to errors on data alone. There is no totally preferred approach. 
The errors from this source are rather small   
-- $\sim 1-5 \%$ except in a few regions of parameter space 
and are similar using various approaches. 
The uncertainty from input assumptions e.g. cuts on 
data, parameterizations {\it etc}., are comparable and sometimes 
larger, which means one cannot entirely believe one group's errors.

The quality of the fit is fairly good, but there are some slight problems.
These imply that errors from higher orders/resummation are 
potentially large in some regions of parameter space,
and due to correlations between partons these affect all regions (the small 
$x$ gluon influences the large $x$ gluon). Cutting out low $x$ and/or $Q^2$ 
data allows a much-improved fit to the remaining data, and altered partons. 
Hence, for some processes theory is probably the dominant source of 
uncertainty at present and a systematic study is a priority as is more data
which would help determine our theoretical accuracy.

\end{document}